\documentclass{jpp}

\usepackage{graphicx}
\usepackage{natbib}
\usepackage{bm}
\usepackage{epsfig}
\usepackage{amsmath}
\usepackage{amsfonts}
\usepackage{amssymb}
\usepackage{epstopdf}
\usepackage{lineno}
\usepackage[dvipsnames]{xcolor}
\usepackage[english]{babel}
\usepackage{float}
\usepackage[utf8]{inputenc}
\usepackage{graphicx}
\usepackage{xcolor}
\usepackage{color}
\usepackage{float}
\usepackage{subfig}

\newsavebox{\astrutbox}
\sbox{\astrutbox}{\rule[-5pt]{0pt}{20pt}}

\newcommand\de{\partial}
\newcommand\pb{$\beta$ }

\newcommand{\8}{\infty}
\def \fig {Fig.~\ref}
\def \eq {Eq.~\ref}
\def \eqs {Eq.s~\ref}

\newcommand{\g}{\textbf}

\newcommand{\rot}{\nabla \times}

\newcommand{\ita}{\textit}
\newcommand{\quotes}[1]{``#1''}
\newcommand{\pan}[1]{(\textit{#1})}

\usepackage{changes}
\definechangesauthor[name={Luca}, color=red]{LSV}
\definechangesauthor[name={Denise}, color=blue]{DP}
\definechangesauthor[name={Oreste}, color=Green]{OP}
\definechangesauthor[name={Sergio}, color=Maroon]{SS}
\definechangesauthor[name={Franco}, color=Gray]{FV}

{\left\lbrace\begin{array}{@{}l@{}}}%
{\end{array}\right.}

\title{Ion Diffusion and Acceleration in Plasma Turbulence}

\author[F. Pecora et al.]{F. Pecora$^1$\thanks{Email address for correspondence: 
francesco.pecora11@unical.it}, S. Servidio$^1$, A. Greco$^1$, W. H. Matthaeus$^2$, D. Burgess$^3$, C. T. Haynes$^3$, V. Carbone$^1$  and P. Veltri$^1$}
\affiliation{
$^1$Dipartimento di Fisica, Universit\`a della Calabria, I-87036 Cosenza, Italy\\
$^2$Bartol Research Institute and Department of Physics and Astronomy, University of Delaware, Newark, DE 19716, USA\\
$^3$School of Physics and Astronomy, Queen Mary, University of London, 327 Mile End Road, London, E1 4NS }

\pubyear{}
\volume{}
\pagerange{}
\date{?; revised ?; accepted ?. - To be entered by editorial office}

\begin{document}
\maketitle
 
\begin{abstract}
Particle transport, acceleration and energisation are phenomena of major importance for both space and laboratory plasmas. Despite years of study, an accurate theoretical description of these effects is still lacking. Validating models with self-consistent, kinetic simulations represents today a new challenge for the description of weakly-collisional, turbulent plasmas. We perform two-dimensional (2D) hybrid-PIC simulations of steady-state turbulence to study the processes of diffusion and acceleration. The chosen plasma parameters allow to span different systems, going from the solar corona to the solar wind, from the Earth's magnetosheath to confinement devices. To describe the ion diffusion, we adapted the Nonlinear Guiding Center (NLGC) theory to the 2D case. Finally, we investigated the local influence of coherent structures on particle energisation and acceleration: current sheets play an important role if the ions Larmor radii are on the order of the current sheets size. This resonance-like process leads to the violation of the magnetic moment conservation, eventually enhancing the velocity-space diffusion. 
\end{abstract}

\begin{PACS}
Authors should not enter PACS codes directly on the manuscript, as these must be chosen during the online submission process and will then be added during the typesetting process (see http://www.aip.org/pacs/ for the full list of PACS codes)
\end{PACS}

\section{Introduction}
\label{sec1}
Processes such as turbulence, diffusion and particle acceleration are ubiquitous both in astrophysical and laboratory plasmas. Understanding particle diffusion is of fundamental importance in order to characterize the distribution of the charged gas in the universe and, more specifically, in the heliosphere. The understanding of the energetic particle motion, originating, for example, from solar flares or coronal mass ejections, can help to prevent injuries for space travelers, as well as hardware damages for satellites. Moreover the distribution of heavy ions in the Earth's magnetosphere can have effects on climate changes \citep{Luo17}. Particle transport theory is fundamental for the dynamics of the solar corona \citep{Lepreti12} and of the interplanetary medium \citep{Ruffolo03,Ruffolo04}. In laboratory plasma experiments, the magnetic confinement could be improved by understanding what affects particle turbulent transport \citep{Taylor71,Hauff09}.

Charged particle dynamics depends on the stochastic motion of the magnetic field lines. The random walk of the magnetic field lines affects the diffusion both across and along the mean magnetic field \citep{JokipiiParker69}. Charged particles gyrate along the magnetic field but, if the field is turbulent, they spread in the perpendicular direction \citep{Ruffolo12,ChandranEA10}, \quotes{jumping} from a field line to another. The turbulent nature of space and laboratory plasmas suggests that the best way to describe their dynamics is given by the statistical approach \citep{Green51,Jokipii66}. In this statistical (Lagrangian) description, particles move in a turbulent electromagnetic field, similarly to the motion of neutral tracers in atmospheric turbulence \citep{Richardson26,Kolmogorov41}. The turbulent nature of the fields that scatter the particles limits the analytical treatment of the subject. For this, we still lack an exact and universal theory for describing the perpendicular diffusion \citep{BAM97,Hussein16}. 

A common theoretical approach, applied when studying diffusion of charged particles, relies on calculating separately the diffusion coefficient in directions parallel and perpendicular to the main guiding field \citep{Jokipii66,Subedi17}. Currently the NLGC theory \citep{Matthaeus03} gives a rather  accurate prediction of the diffusion coefficient for systems with a three-dimensional (3D) geometry. However, this theory has been tested only with test-particle simulations - simulations that do not take into account the back-reaction of the particles motion on the electromagnetic fields. Nowadays, more realistic simulations of turbulence are available and is therefore interesting to test the validity of these predictions, using, for example, kinetic models.

A full 3D description of plasma turbulence requires huge computational efforts that can be lightened by reducing the dimensionality of the problem. Magnetohydrodynamics (MHD) turbulence simulations \citep{DobrowolnyEA80,Shebalin83,Dmitruk04} have shown that, if a strong guiding magnetic field is present, turbulence loses the property of isotropy. The magnetic structures are mainly present in the plane perpendicular to the main field \citep{BrunoCarbone16}. These structures are essentially composed by a sea of flux tubes (magnetic islands) and current sheets, where reconnection might eventually occur \citep{Matthaeus86,Greco09,Servidio11}. In this complex pattern, particles diffuse and accelerate. 

MHD simulations suggest that turbulence can be considered as a network of reconnecting magnetic islands, that merge, change topology and convert energy \citep{Parker57,Matthaeus84,Ambrosiano88,ServidioEA09}. Magnetic reconnection is considered one of the most effective mechanisms for particle acceleration and energisation \citep{Zank14}, being crucial for explosive events in the solar atmosphere, like solar flares \citep{Cargill06,Cargill12,Nigro14} and coronal mass ejections \citep{Gosling10}. Analogously, in the Earth's magnetosphere \citep{Drake06,Oka10,Birn12} and far away to the heliopause \citep{Lazarian09}, magnetic reconnection is thought to be a very active mechanism. Observations and measurements strongly relate the magnetic island merging \citep{Khabarova14} and the magnetic field discontinuities \citep{Tessein13} with the increase of suprathermal energetic particles. This increasing number of energetic particles in zones of merging magnetic islands and of strong discontinuities can be the key interpretation to relate the magnetic reconnection with acceleration and energisation \citep{LeRoux15,Lazarian09}.

Both 2D and 3D simulations of MHD turbulence with a strong guiding magnetic field \citep{Matthaeus86,Gray92,Dmitruk04} reveal power law energy spectra for test-particles. The power-law spectra can be indeed observed for anomalous cosmic rays \citep{Stone08,Decker10}, energetic ions in the solar wind \citep{Fisk06} and energetic electrons in solar flares \citep{Holman03}.  It is worth noting, however, that since test-particle models are not self-consistent, particles in these simulations can reach extremely high energies. These very high energies might not be commonly observed in the solar wind, in usual turbulence conditions, especially for ions, suggesting that a self-consistent kinetic model need to be used in order to correctly describe the plasma dynamics.

The present work is organized as follows. In Section \ref{sec2} we will present a global overview on the model that describes the plasmas in the low-collisionality limit. We will introduce the numerical algorithm that simulate the plasma-particle dynamics. In Section \ref{sec3}, the particle trajectory statistics will be presented, interpreting the numerical results via the plasma turbulent diffusion theories. The investigation of the acceleration process will be the central topic of Section \ref{sec4}, where we will perform local analysis of the acceleration phenomenon.  Finally, in the last Section, we will present the discussions, conclusions and future perspectives.

\section{Simulations of 2D Plasma Turbulence}
\label{sec2}
We perform three hybrid-PIC simulations, varying the plasma $\beta$ (the ratio between kinetic and magnetic pressure), with $\beta=5$, $0.5$ and $0.1$, in order to cover a wide range of relevant plasma scenarios. These chosen values are such that the high \pb value is typical of plasmas found in the Earth's magnetosheath; the value of $\beta=0.5$ is close to the typical solar wind conditions, whereas the lowest \pb is appropriate for the the solar corona and laboratory devices.

Particles and fields are described by the Vlasov-Maxwell system \eqs{eq:hybridPIC} that we solve via a hybrid-PIC approach, using kinetic ions and fluid electrons. The system of equations is given by

\begin{align}
\begin{aligned}
&\dot{\g{x}} = {\bf v}\\
&\dot{\g{v}} = {\bf E}+{\bf v}\times{\bf B}\\
&\dfrac{\de \g{B}}{\de t} = - \rot \g{E}\\
&\g{E} = - ( \g{u} \times \g{B} ) + \dfrac{1}{n} \; \g{j}\times\g{B} - \frac{1}{n}\nabla P_e +\eta \g{j} .
\end{aligned}
\label{eq:hybridPIC}
\end{align}

In the above equations, $\g{x}$ is the particles position, $\g{v}$ their velocity, $\g{E}$ is the electric field, $\g{B}$ is the magnetic field, $\g{u}$ is the proton bulk velocity (the first moment of the velocity distribution function), $n$ is the proton number density (the zero-th moment of the ion velocity distribution function), $\g{j}$ is the current density. The pressure term is adiabatic $P_e = \beta n^\gamma$ and $\eta=0.006$ is the resistivity that introduces a small scale dissipation, for numerical stability. The electric field is given by the generalised Ohm's law. In the simulations distances are normalized to $c/\omega_{p_i}$, where $c$ is the speed of light and $\omega_{p_i}$ is ion plasma frequency. The time is normalized to $\Omega_{c_i}^{-1}$, that is the ion cyclotron frequency. Finally, velocities are normalized to the Alfvén speed $v_{{}_A}= c \Omega_{c_i}/\omega_{p_i}$.

The three simulations have the same initial conditions: uniform density and a Maxwellian distribution of particles velocities with uniform temperature. We impose large scale fluctuations in order to mimic the motion of the large energy-containing vortices. We have chosen random fluctuations, at large scale, for both magnetic field and the ion bulk velocity field. The \eqs{eq:hybridPIC} are solved on a square grid of size $L_0=128 d_p$, where $d_p$ is the proton skin depth defined as $d_p = c /\Omega_{c_i}$, discretized with $512^2$ points, with periodic boundary conditions. This initial state consists of a 2D spectrum of fluctuations, perpendicular to the main field $B_0$ (the latter chosen along $z$). The fluctuations amplitude is $\delta b/B_0 \sim 0.3$. To suppress the statistical noise of the PIC method we use $1500$ particles per cell (about $4\times 10^8$ total particles).

To describe the magnetic topology, in the 2.5D approximation, it is useful to define the in-plane magnetic field as $\g{B}_\perp = \nabla a_z \times \hat{\bf z}$, where $a_z$ is the magnetic potential and $\hat{z}$ is the out-of-plane (axial) unit vector. The current in the axial direction $j_z = (\rot \g{B}_\perp ) \cdot \hat{z} = -\nabla^2 a_z$. To achieve a stationary state of fully developed turbulence, we initially let the system decay freely, and then we introduce a forcing at the time $t^*$ at which nonlinearity reaches its peak (roughly the peak of $\langle j_z^2\rangle$), namely $t^* \sim 25 \Omega_{cp}^{-1}$. The forcing consists of ``freezing'' the amplitude of the large-scale modes of the in-plane magnetic field, with $1\leq m\leq 4$, with constant phases. This corresponds to a large-scale input of energy, as described in \citet{Servidio16}.

\begin{figure}
\centering
\subfloat
{ \includegraphics[width = \textwidth]{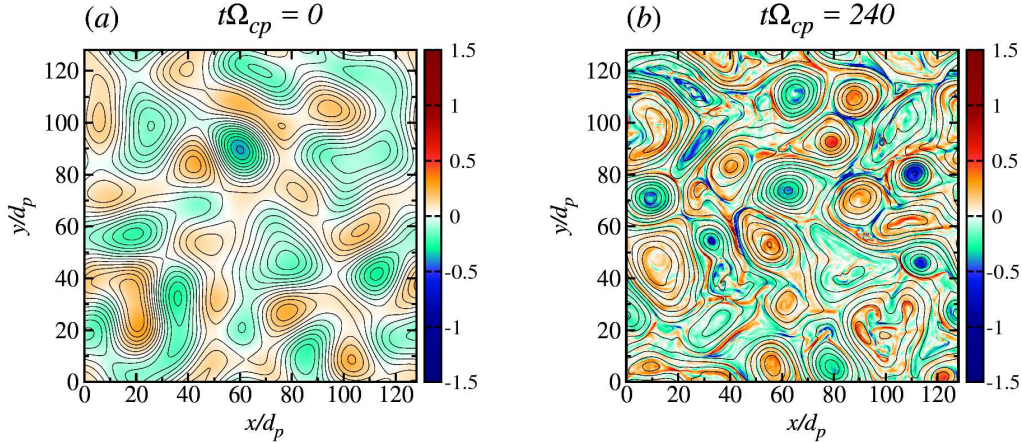} }
\caption{Current density $j_z$ (colour map) together with vector potential $a_z$ (contour lines), at the initial state \pan{a} , with big structures and tenuous axial current, and at the final state \pan{b}, with coherent structures such as small vortices and intense current sheets. The most intense current sheets are located in between reconnecting magnetic islands \citep{MattMon80}.}
\label{fig:jz}
\end{figure}

In order to have a significant statistics, we perform our analysis when a steady state has been achieved, namely for $50<t\Omega_{cp}<250$. \fig{fig:jz} shows the shaded contour of the current density $j_z$ along with the contour line of the vector potential $a_z$, as the system evolves toward turbulence. Panel \pan{a} shows the initial state of the system, where islands and current sheets are not defined yet. As time goes on, smaller vortices and sharp current sheets develop. In particular, these magnetic structures represent magnetic islands (flux tubes in 3D). The global appearance of the system remains unchanged, in a statistical sense, when the peak of nonlinearity has been reached. The regions of big magnetic gradient appear in between reconnecting magnetic islands, with associated intense current sheets \citep{MattMon80,Servidio15}.

\begin{figure}
\centering
\subfloat
{ \includegraphics[width = \textwidth]{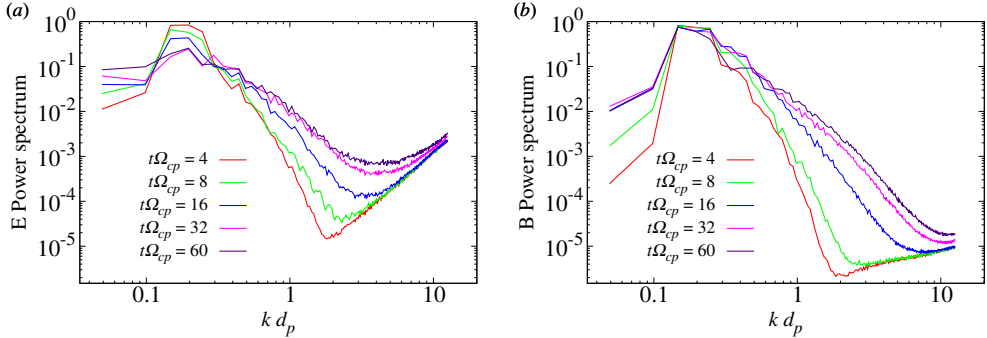} }
\caption{Power spectra of electric \pan{a} and magnetic \pan{b} fields as a function of the wavenumber (normalized with the proton skin depth $d_p$), at different times. While turbulence develops, energy is transferred from the large to the small scales. In the steady state, at $k\sim d_p^{-1}$, the electric field spectrum manifests a steeper slope.}
\label{fig:EB_dev_spectra}
\end{figure}

To better describe the state of fully developed turbulence, we computed the Fourier spectra, as a function of the wavenumber $|\g{k}|$, of both the electric and magnetic fields. The power spectra for the trace of the correlation tensor, $|\tilde{\g{B}}({\g{k}, t})|^2$, where $\tilde{\g{B}}({\g{k}, t})$ are the respective Fourier coefficients, are reported in \fig{fig:EB_dev_spectra}. These spectra exhibit the classical scenario of MHD turbulence, in which energy flows from large to small scales. This cascade of energy occurs over the so-called inertial subrange, where energy that scales as $k^{-5/3}$ \citep{BrunoCarbone16}. In particular, \fig{fig:EB_dev_spectra} shows the turbulent development. The state of fully developed turbulence is achieved after $50\Omega_{cp}^{-1}$. After this time the spectra are almost stationary (do not experience large fluctuations). It is important to note, that as observed in the solar wind \citep{Bale05}, the power in the electric fields is higher, at $k$'s that correspond to characteristic ion lengths. This is in agreement with previous studies and simulations of plasma turbulence \citep{howes08a,MatthaeusEA08,FranciEA15}.

In order to understand the dynamics of particles in a turbulent scenario, where structures are present at different scales, it is useful to measure different characteristic lengths. The correlation (or integral) length of a turbulent field generally corresponds approximately to the size of the largest energy-containing turbulent eddies, (in our case the typical sizes of the islands.) This length can be obtained from the two-point correlation function $C(\g{r})$ as
\begin{equation}
\label{eq:lambdac}
\lambda_{{}_C} = \int_\Omega C(\g{r}) d\g{r} =   \dfrac{1}{\langle b^2 \rangle}\int_\Omega d\g{r} \langle \g{b}(\g{x}+\g{r})\cdot \g{b}(\g{x}) \rangle.
\end{equation}
Here $\g{b}$ represents the magnetic fluctuations 
and $\langle \cdot \rangle$ is the average over the total volume $\Omega$. Note that fluctuations are isotropic in the plane perpendicular to the main field, hence $\g{x}$ and $\g{r}$ are in-plane vectors and the above length is essentially the same, namely $C(r_x)\sim C(r_y)\equiv C(\g{r})$. Moreover, $\lambda_C$ is very similar for the bulk velocity fluctuations, because of the choice of the initial conditions and the driving. In the above definitions we suppressed the time dependence because  of stationary, and we computed these lengths averaging over time, in the steady state regime. For our system $\lambda_C \sim 10 d_p$ and does not vary for runs wit different plasma $\beta$.

Apart of the above energy-containing scale, which is a large scale characteristic length of turbulence, it is important to characterize also the smallest scales properties. In hydrodynamics the Taylor length is the scale at which the viscous dissipation term is no longer negligible \citep{Servidio11}. This scale is related to the largest width of the structures,  where dissipation starts to be relevant. In our case, the magnetic Taylor scale  can be defined as
\begin{equation}
\label{eq:lambdat}
\lambda_{{}_T} = \sqrt{\dfrac{b_\perp^2}{\langle j_z^2 \rangle}}.
\end{equation}
In the above expression, $b_\perp$ is the root mean square of the magnetic field in the plane perpendicular to the axial direction (perpendicular to the main guiding field), and $\langle j_z^2 \rangle$ is the axial current averaged over the 2D space, and averaged also over the times in which the system is in the stationary state of fully developed turbulence. This scale can be considered as the scale at which the inertial range of turbulence terminates \citep{Frisch95}. In our simulations $\lambda_{{}_T}\sim 1.7 d_p$. Since the Taylor length $\lambda_{{}_T}$ gives the biggest size of the dissipative structures, we introduce now an average measure of these structures, namely the typical width of the layer cores (O-points). The current sheet width, $\delta_c$,  is the half maximum width of the current sheet intensity profile. The average core width of the current sheets we measured as $\delta_c \sim 0.3 d_p$.

\section{Particle diffusion in plasma turbulence}
\label{sec3}

\begin{figure}
\centering
\subfloat
{ \includegraphics[width=\textwidth]{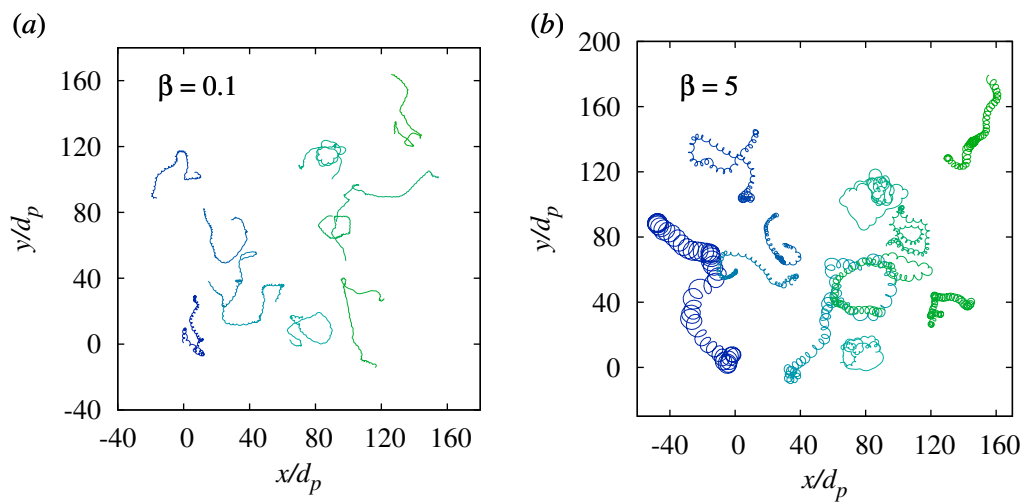} }
\caption{Particle trajectories for different values of $\beta$, namely $\beta = 0.1$ \pan{a} and $\beta=5$ \pan{b}. It is possible to observe that particles in high \pb have bigger Larmor radii, while particles in low \pb one move almost on straight lines, with more abrupt change of direction. One can notice some particles have closed orbits like they are trapped in a magnetic island while others experience turnovers that suddenly bend the trajectory.}
\label{fig:2dtraj}
\end{figure}

We investigate the particle motion in steady state turbulence. Among all PIC macro-particles, we have followed the path of $10^5$ samples, verifying the convergence of the statistical results. From their  positions as a function of time, it is possible to compute the diffusion coefficient, measuring their mean squared displacement. After a general overview on the erratic trajectory of particles in 2D turbulence, we will see whether the plasma $\beta$ affects the statistics of ions diffusion and acceleration.

In \fig{fig:2dtraj} we show the 2D trajectories of some randomly selected particles, during the whole simulation, for two distinct values of $\beta$ (low and high $\beta$). As it can be seen, particles that move in high \pb plasma are less magnetized and their Larmor radii is large. Particles in low-\pb plasma, instead, are highly magnetized and perform tight gyrations, with more abrupt change of direction. Small-$\beta$ particles spread less than in the high $\beta$ case. Generally, in all cases, one can notice that some particles have closed orbits as they are trapped in a magnetic island while others experience turnovers that suddenly bend the trajectory. As we shall see, particles that perform closed trajectories are mostly trapped in turbulent vortices, whereas sharply segmented trajectories can belong to particles which encounter local intense turbulent structures, as magnetic discontinuities or current sheets, that make them deviate suddenly \citep{DrakeEA10,HaynesEA14}.

\begin{figure}  
\centering

\includegraphics[width=0.7\textwidth]{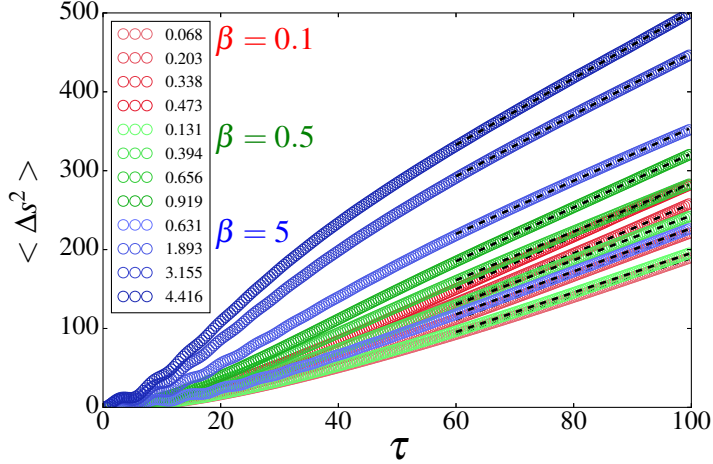}
\caption{Mean squared displacement in the perpendicular plane as a function of the time $\tau$ (cyclotron units), for several particles energy  ranges. For small times, it is possible to observe a deviation from the linear prediction of the Brownian motion (\eq{eq:brown_disp}). Normal diffusion is achieved at later times, for $\tau>60 \Omega_{cp}^{-1}$, where we extrapolate the values of $D$ via linear fits (dashed lines).}
\label{fig:brown_2dt}
\end{figure}

The overall motion, as represented in \fig{fig:2dtraj}, seems quite random: particles scatter and diffuse in time. We computed the square displacement $\Delta s^2=\Delta x^2+\Delta y^2$, finding that, for sufficiently long time intervals, the motion is diffusive, namely
\begin{equation}
\langle \Delta s ^2 \rangle = 2D \tau.
\label{eq:brown_disp}
\end{equation}
In the above expression, $D$ is the diffusion coefficient, $\tau$ is the time interval over which the particle moves by $\Delta s$, and the brackets now represent an ensemble average computed over particles.
Note that $\tau=t-t_0$, and because of stationarity we varied $t_0$ in the range of the steady state regime. To understand the dependence of the diffusion coefficient on particle energy, we divide the particles in energy bins and evaluate the above diffusion coefficient, collecting particles depending their average energy. The averaged mean squared displacement as a function of the time interval for several energy values is shown in \fig{fig:brown_2dt}. It is possible to observe the normal diffusive behavior is achieved for for $\tau \gtrsim 60 \Omega_{cp}^{-1}$. By fitting the curves after this interval with the \eq{eq:brown_disp}, we can obtain the numerical value of $D$.
For shorter time intervals particles seems to not follow the linear law. Indeed, for short time intervals the particle motion cannot be stochastic since they have ``memory'' of their initial condition, and temporal correlations exist.  It is clear that the higher is the particles energy, the larger is the diffusion coefficient. Before reporting the numerical results on the diffusion of particles in self-consistent kinetic simulations, we will introduce the theoretical background.

\subsection{A Non Linear Guiding Center (NLGC) model in 2D}
\begin{figure}
\centering
\subfloat
{ \includegraphics[width=0.5\textwidth]{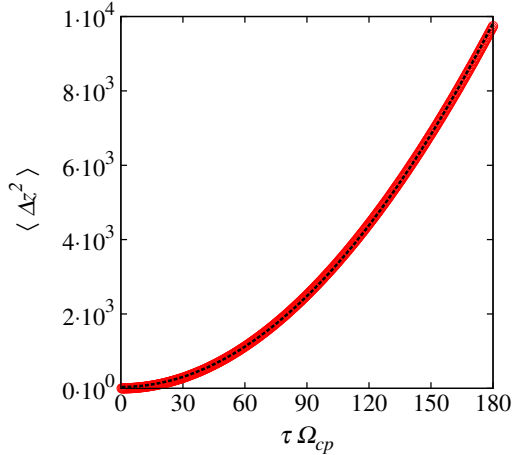} }
\caption{Mean squared displacement in the $z$ direction averaged over all particles as a function of $\tau$. The dashed line is the parabolic fit ($\propto \tau^2$). For every $\beta$, the assumption of free streaming made in \eq{eq:vz2appr} is pretty solid.}
\label{fig:cvz}
\end{figure}

We want to develop a theory that can describe particle diffusion in 2D turbulence, and test it with the self-consistent plasma simulations. Currently, a useful description of perpendicular diffusion is given by the Non Linear Guiding Center (NLGC) theory \citep{Matthaeus03}. This theory works well for 3D turbulence, with a uniform mean magnetic field, and it has been tested with test-particle simulations. To develop the 2D version of the theory we start from the 3D NLGC derivation, considering the particle gyromotion. Using the Taylor-Green-Kubo (TGK) formulation \citep{Taylor22, Green51, Kubo57, ShalchiEA08, Shalchi15}, one can write the diffusion coefficient $D_{xx}$, for instance along the $x$ direction, as
$$
D_{xx} = \dfrac{1}{B_0^2}  \int_0^{\infty} d\tau \langle v_z(0)B_x(\g{x}(0),0) v_z(\tau)B_x(\g{x}(\tau),\tau) \rangle,
$$
where it is assumed that the particle motion projected on the 2D plane follows the 2D  projection of the magnetic field lines.
We assume that the magnetic field fluctuations in the perpendicular plane are completely uncorrelated with the velocity in the $z$ direction and this allows us to write the diffusion coefficient as
$$
D_{xx} = \dfrac{1}{B_0^2}  \int_0^{\infty} d\tau \langle v_z(0) v_z(\tau) \rangle \langle B_x(\g{x}(0),0) B_x(\g{x}(\tau),\tau) \rangle.
$$
Now we encounter a deviation from the 3D NLGC, by saying that the velocity correlation function in the $z$ direction is nothing but the square value of the $z$ velocity i.e.
\begin{equation}
\langle v_z(0) v_z(\tau) \rangle \sim  v_z^2.
\label{eq:vz2appr}
\end{equation}
This assumption has been directly proved by measuring the quantity $\langle v_z(0) v_z(\tau) \rangle$ in our simulations. \fig{fig:cvz} shows that the motion along $z$ is given by free streaming, with $\langle \Delta z^2 \rangle \sim t^2$ and therefore the velocity $v_z$, in the direction along the main field, is almost constant.

Using this approximation, we can estimate the diffusion coefficient neglecting the parallel scattering, and get: 
$$
D_{xx} = \dfrac{v_z^2}{B_0^2}  \int_0^{\infty} d\tau \; \langle B_x(\g{x}(0),0) B_x(\g{x}(\tau),\tau) \rangle.
$$
Using the Corrsin's independence hypothesis to transform the integrand function as
$$
\langle B_x(\g{x}(0),0) B_x(\g{x}(\tau),\tau) \rangle = \int d\g{r} \; R_{xx}(\g{r},\tau) P(\g{r},\tau)
$$
one arrives at
$$
D_{xx} = \dfrac{v_z^2}{B_0^2}  \int_0^{\infty} d\tau \; \int d\g{r} \; R_{xx}(\g{r},\tau) P(\g{r},\tau), 
$$
where $R_{xx}(\g{r},\tau)$ is the Eulerian two-point two-time correlation tensor and $P(\g{r},\tau)$ is the probability function of the particle having displacement $\g{r}$ after a time interval $\tau$ \citep{Matthaeus03}. At this point is convenient to express $R_{xx}(\g{r},\tau)$ using its Fourier transform
$$
R_{xx}(\g{r},\tau) = \int d\g{k} \; S_{xx}(\g{k},\tau) e^{i\g{k}\cdot\g{r}}.
$$
We can model $S_{xx}(\g{k},\tau) = S_{xx}(\g{k})\Gamma(\g{k},\tau)$, where $\Gamma(\g{k},\tau)$ represents the time-propagator of the spectrum, and $S_{xx}(\g{k})$ is spatial spectrum. Usually, this functional form is described via the so-called sweeping decorrelation mechanism, namely $\Gamma(\g{k},\tau) \sim e^{-\tau/\tau_c(\g{k})}$, where $\tau_c(\g{k})$ is a characteristic decorrelation time. The latter is usually thought to be the  the ``sweeping decorrelation time'' \citep{ChenKraichnan89,NelkinTabor90}, in analogy with fluid turbulence (see below). 

At this point, we can proceed with the theoretical estimate of the diffusion coefficient, given by
\begin{equation}
D_{xx} = \dfrac{v_z^2}{B_0^2} \int_0^\8 \int d\g{k} \; S_{xx}(\g{k}) d\tau \Gamma(\g{k},\tau)  \int d\g{r} \; P(\g{r}, \tau) e^{i\g{k}\cdot\g{r}}.
\end{equation}
Since $P$ is Gaussian, then
$$
\int d\g{r} \; P(\g{r}, \tau) e^{i\g{k}\cdot\g{r}} = e^{-(k_x^2D_{xx} + k_y^2D_{yy})\tau}.
$$
In In axisymmetric turbulence $D_{xx} = D_{yy} \equiv D$, $S_{xx} = S_{yy} \equiv S$ and, since we have only in-plane structures (i.e. $k_z=0$) $k_x^2 + k_y^2 \equiv k^2$, the diffusion coefficient is
$$
D = \dfrac{v_z^2}{B_0^2}  \int d\g{k}\; S(\g{k}) \int_0^\8 d\tau \; e^{-k^2D\tau} e^{-\tau/\tau_c({\bf k})}.
$$

Finally, integrating over $\tau$, one gets
\begin{equation}
D = \dfrac{v_z^2}{B_0^2} \int d\g{k} \; \dfrac{S(\g{k})}{\left[\tau_C(k)\right]^{-1} + k^2D}.
\label{eq:2DNLGCD}
\end{equation}
Note that this prediction is only a small modification to the NLGC theory, having suppressed the $z$-dependence by using the particle free streaming along $z$ (\eq{eq:vz2appr}). 

In order to obtain a first estimate of the diffusion coefficient $D$, we can make an approximation.  Assuming that the sweeping decorrelation time is rather long, (as in the case of large scale slow driving, for example) we can drop the term $\tau_C(k)^{-1}$ in \eq{eq:2DNLGCD}. This latter term, indeed, makes the equation more complex to solve. We therefore obtained:
\begin{equation}
D^* \sim \sqrt{ \dfrac{v_z^2}{B_0^2}\int d\g{k} \dfrac{S(\g{k})}{k^2} }.
\label{eq:apprD}
\end{equation}
The reader may notice that this result is essentially of the same form as the so-called Field Line Random Walk (FLRW)  limit of perpendicular particle scattering \citep{Jokipii66,BAM97}. For such cases, $D= v_{eff} D_{fl}$ where $D_{fl}$ is the Fokker Planck coefficient for field line transport, and $v_{eff} $ is the effective velocity of the particle along the magnetic field \citep{Matth95}.

\begin{figure}
\centering
\subfloat
{ \includegraphics[width=0.8\textwidth]{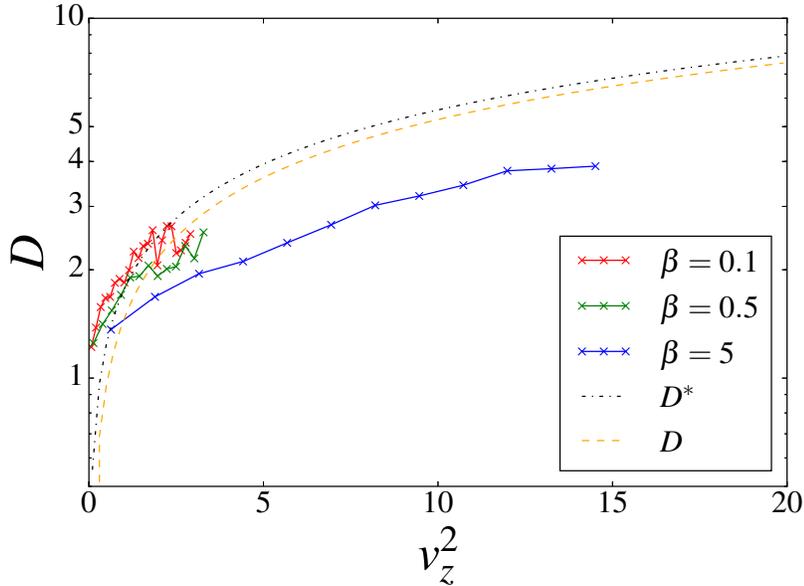}}
\caption{The diffusion coefficient as a function of the parallel energy $v_z^2$, calculated by fitting the mean square displacement (red, green and blue line-points respectively for $\beta = 0.1$, $0.5$ and $5$) for all the simulations.
The  theoretical diffusion coefficient from the approximated 2D NLGC theory (\eq{eq:apprD}) is reported with the black point dashed line whereas the exact calculation (\eq{eq:2DNLGCD}) is the orange dashed line.}
\label{fig:D_v22}
\end{figure}

To evaluate the diffusion coefficient via \eq{eq:apprD} and \eq{eq:2DNLGCD}, we computed the average power spectrum $S(\g{k})$, from the simulations. \fig{fig:D_v22} shows the numerical values of $D$ obtained by fitting the particles mean squared displacements in \fig{fig:brown_2dt}. The theoretical values of the 2D NLGC, evaluated via both the exact and the approximated formulas, are reported as a function of the particles energies. The simulations results follow fairly well the theoretical prediction at low \pb. The theory slightly deviates at very high energy ($\beta$'s), although the functional monotonic behaviour is very similar. We will note that we do not observe large energy excursions like in test-particle models, due to the self-consistency of the ions treatment.

\begin{figure}
\centering
{\includegraphics[width=\textwidth]{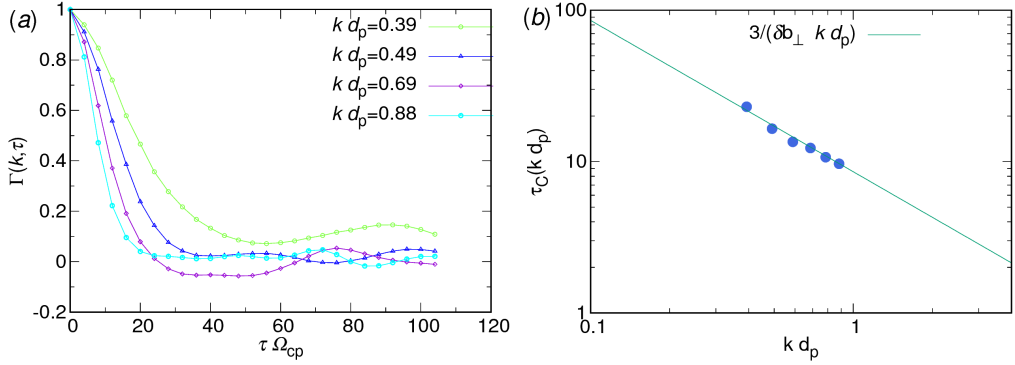} }
\caption{\pan{a} Propagator of the magnetic field spectrum, computed for several (perpendicular) $k$ modes in the inertial range of plasma turbulence. \pan{b} Decorrelation time as a function of $k$ (blue bullets), computed as the $e$-folding time of the functional form in panel \pan{a}. The sweeping prediction is reported with dashed (green) lines.
}
\label{fig:rab}
\end{figure}

Here we briefly investigate the decorrelation mechanism in self-consistent, plasma turbulence. We performed a Fourier transform in time of the magnetic fluctuations, computing the propagator $\Gamma(k, \tau)$, as described in Ref.s \citep{ServidioEAL11,PerriEA17}. As in fluid, MHD, and Hall MHD models of turbulence, this time-dependent correlation of turbulence strongly depends on the amplitude of $k$, as reported in \fig{fig:rab}. As it can be seen, the decorrelation mechanism depends on $k$ and drops quickly in time (only a few inertial range modes are reported). From this functional form, we computed the decorrelation time $\tau_C(k)$, represented in the panel \pan{b} of the same figure. The decorrelation time scales as $\sim 1/k$, indicating the clear dominance of the sweeping effect. To be more quantitative, in order to compute the diffusion coefficient for our experiment, we found that $\tau_C(k) \sim 3/(\delta b_\perp k d_p)$, where $\delta b_\perp$ is the $rms$ of the in-plane magnetic fluctuations.

\section{Particle Acceleration}
\label{sec4}
In agreement with previous literature \citep{JokipiiParker69, Matthaeus03}, we have found that diffusivity depends on particles energy. The behavior of diffusion is in agreement with a modified model of the NLGC theory, here adapted to collisionless 2D plasmas. Moreover, we observed that many of these particles experience variations of their momentum -- namely an acceleration process. In this section, we investigate the mechanisms responsible for acceleration, using conditional statistics. We will establish the relationship between the effectiveness of the acceleration mechanism and the particle energies, and the existence of possible resonance conditions that energize the ions.

\begin{figure}
\centering
\includegraphics[height=0.5\textwidth]{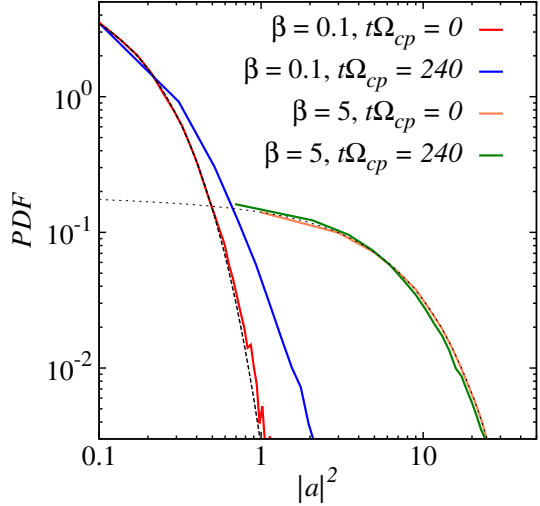}
\caption{PDFs of the acceleration, for $\beta = 0.1$ and $\beta= 5$. Both initial and final times of the simulations are reported, together with the corresponding probability distributions (dashed black lines). In the low beta case, extreme acceleration processes are observed.
}
\label{fig:PDFa2}
\end{figure}

We have computed, for each particle, the Lagrangian acceleration ${\g{a}}= {\partial \g{v}}/{\partial t}$,  where $\g{v}$ is the particle velocity. We computed the the Probability Distribution Function (PDF) of the Lagrangian acceleration, for all the particles, at different times and for different simulations. The acceleration has been computed by using a 6$^{th}$ order finite difference (centered) method. The distributions are reported in \fig{fig:PDFa2}, for the two extreme $\beta$, at the beginning and at the end of the simulation. As it can be observed, the PDFs of particle acceleration in high \pb plasma are well described by the $\chi^2$ distribution. Let $\lbrace x_1, x_2, \dots, x_k \rbrace$ be a set of $k$ independent normally distributed variables. The sum of the square of these $k$ variables distributes according to the $\chi^2_k$ PDF that is defined as
$$
PDF(\chi^2_k) = \dfrac{1}{2^k \Gamma(k/2)} x^{k/2-1} e^{-x/2}
$$
where $k$ are commonly referred to as degrees of freedom, and $\Gamma(x)$ is the gamma function. In our case the three independent variables are the acceleration components $\lbrace a_x, a_y, a_z \rbrace$ and the square modulus of the acceleration $|a|^2 = a_x^2+a_y^2+a_z^2$ distributes according to the $\chi^2_3$ with 3 degrees of freedom.

For $\beta=5$,  in particular, this distribution does not change in time, indicating the lack of very extreme events during the evolution of the system. Whereas at low $\beta$ ($\beta = 0.1$), the acceleration, which is initially randomly distributed, develops a tail at later times (blue solid line). This means that the number of particles with anomalous acceleration is higher for low \pb plasma, suggesting that a physical process that depends on the plasma conditions is at work.

\begin{figure}
\centering
\subfloat
{ \includegraphics[width=0.8\textwidth]{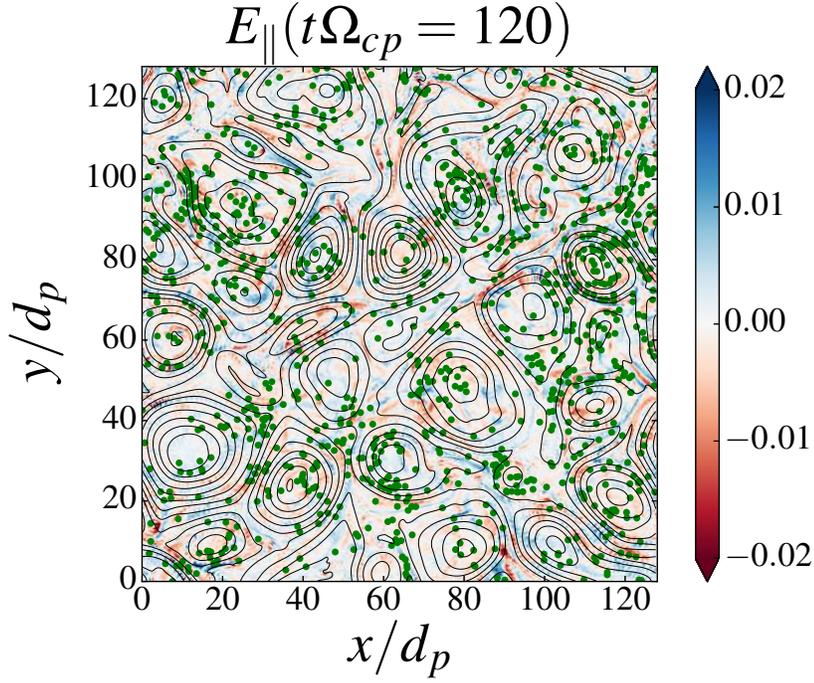} }
\caption{Particles with very high acceleration (green dots), superimposed on the shaded contour of the parallel electric field. These particles are localized in the regions where the parallel electric field is bigger.}
\label{fig:aQmap}
\end{figure}

A straightforward acceleration mechanism can be due to an electric field parallel to the local magnetic field: 
\begin{equation}
E_\parallel = \dfrac{\g{E}\cdot\g{B}}{|\g{B}|}.
\end{equation}
We identified the anomalous particles, i.e. the particles with acceleration values exceeding the variance of the followed distribution and represented their positions, at a given time, over the parallel electric field map with a contour plot of the magnetic potential in the $z$ direction, in order to see whether they show a correlation with the turbulent structures. This map is reported in \fig{fig:aQmap}. The figure shows that particles with anomalous acceleration are affected by the above field: accelerating ions are non uniformly distributed, they cluster where the parallel electric field is more intense, on the flanks of the magnetic islands.

\begin{figure}
\centering
{ \includegraphics[width=\textwidth]{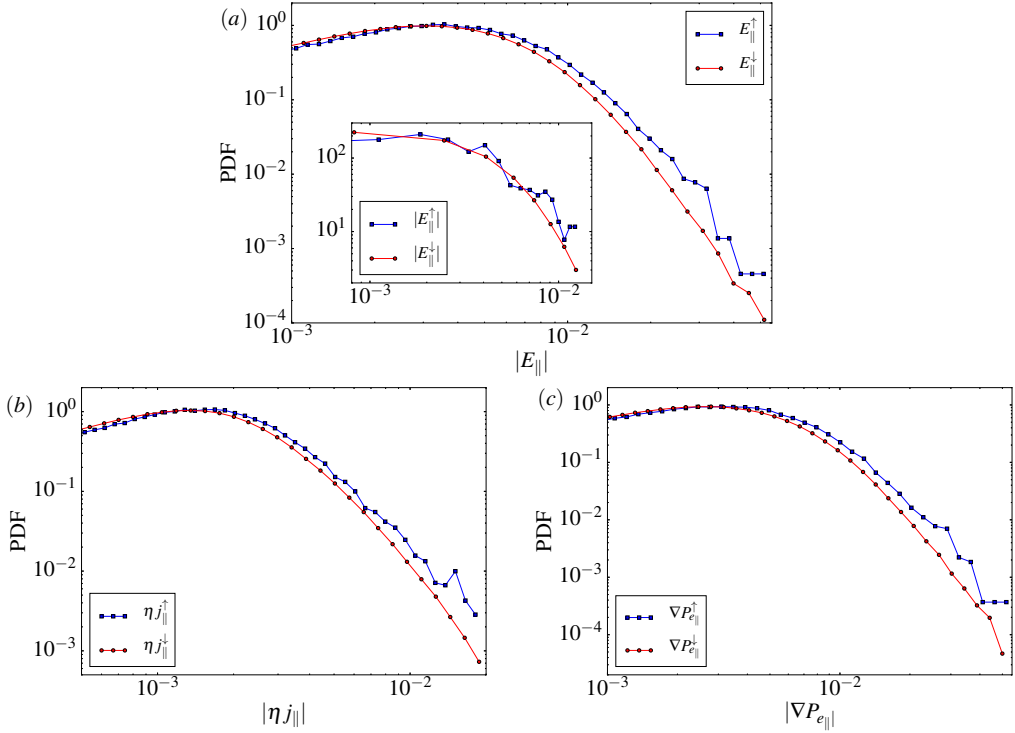} }
\caption{Conditioned PDFs of the electric field $E_\parallel$ \pan{a} and of its single components $\eta j$ \pan{b} and $\nabla P_e$ \pan{c} in the direction parallel to the local magnetic field. The condition is related to the particles acceleration values. In particular, the fields with superscript ${}^\uparrow$ (blue dots) are measured at the position of particles with acceleration exceeding $3\sigma$, whereas the fields labeled with ${}^\downarrow$ (red dots) are measured where particles with acceleration less than $1\sigma$ lie. The whole parallel electric field and both its component show the same behaviour: anomalous particles are more likely to be found where the field values are bigger. This indicates that particles cluster close to regions where dynamical activity is occurring near current sheets, suggesting an association with magnetic reconnection.  
The inset of panel \pan{a} shows the total parallel electric field component conditioned over the PVI values. The red circles are the parallel electric field values computed along the whole PVI path, whereas the blue squares are the parallel electric field values computed in the regions where the PVI exceeds a threshold value. This correlation indicates that high electric field values are more probable to be found where magnetic field inhomogeneities are stronger.
}
\label{fig:QPDF}
\end{figure}

In order to establish a more quantitative link between the parallel electric field and possible local acceleration effects, we compared the PDFs of the single terms of the electric field that can be parallel to the local magnetic field (namely $\nabla P_e$ and $\eta j$) and of the total parallel electric field itself ($E_{||}$) 

over the positions of non-anomalous particles, and the PDFs of the same quantities measured only at the positions of high-acceleration particles. The particles were distinguished by thresholds with respect to the global acceleration PDF: the anomalous particles are those which have an acceleration value exceeding the $3\sigma$ of the global distribution, whereas \quotes{normal} particles have an acceleration value within $1\sigma$. Note that the above conditional statistics would give the same distribution only if acceleration and $E_{||}$ are uncorrelated. These PDFs are shown in \fig{fig:QPDF}, where the quantities with superscript ${}^\uparrow$ are measured at the positions of anomalously accelerated ions, whereas the fields with superscript ${}^\downarrow$ are related to \quotes{normal} particles. As it can be seen, accelerated particles have higher parallel electric field --  the population of the distribution is higher at big $E_{||}$ regions. Moreover we used the Partial Variance of Increments (PVI) method to find candidate regions likely to be identified as coherent structures. These structures contribute to non-Gaussian statistics and therefore to intermittency. The PVI time series is substantially defined as the normalized sequence of magnetic increments. By applying the PVI technique, we found that the regions of bigger $E_{||}$ occur in correspondence of magnetic discontinuities and not in smooth regions, as the PDF of $E_{||}$ conditioned on PVI values \citep{Greco09PRE} clearly evidences in the inset of panel \pan{a} of \fig{fig:QPDF}.

This statistics further confirm the relation between the parallel electric field and the stochastic acceleration mechanisms in 2D turbulence.

\begin{figure}
\centering
{ \includegraphics[width=\textwidth]{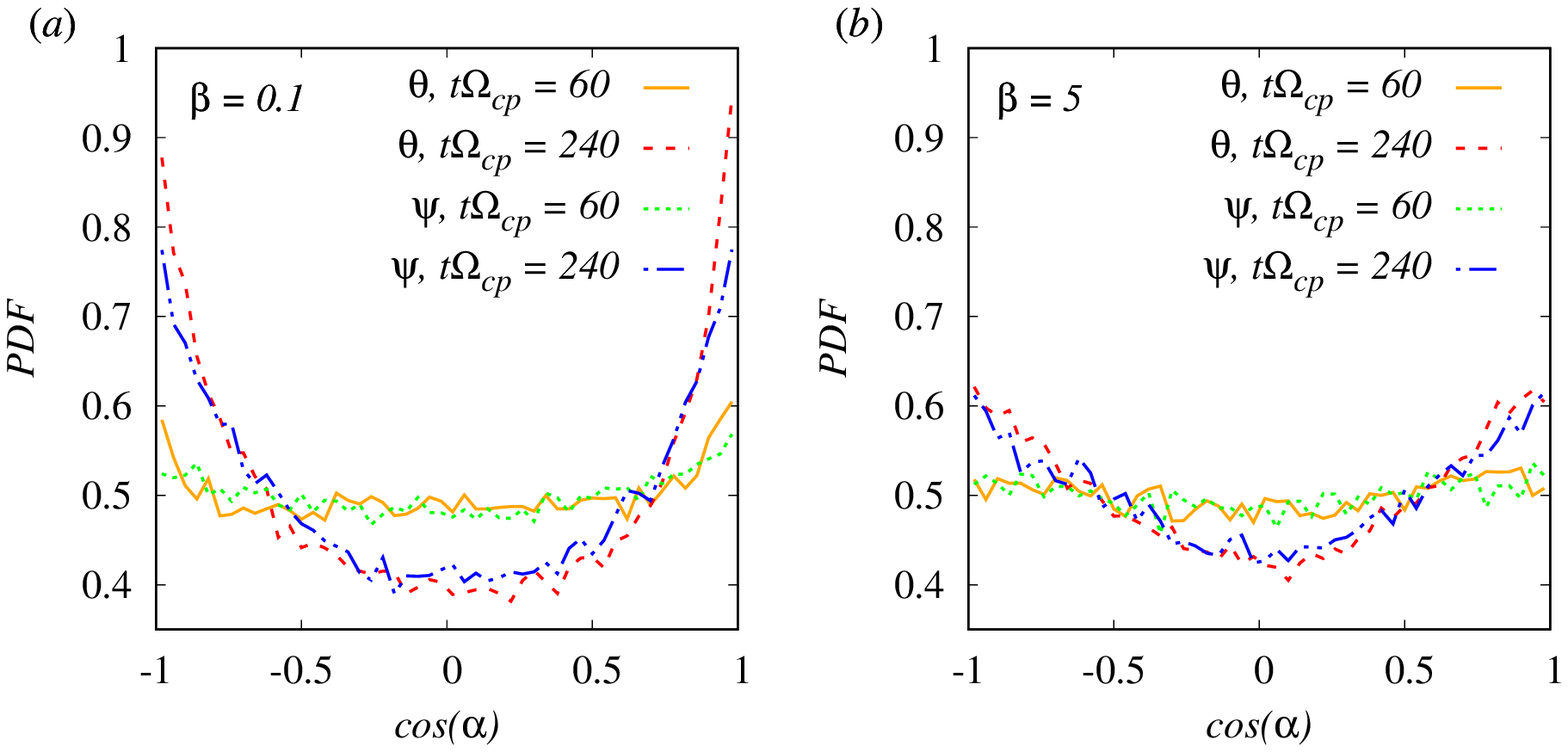} }
\caption{PDFs of the cosine of the angle between the particles velocity and the magnetic field for low \pan{a} and high \pan{b} \pb values at the initial and final instants of the simulation. The orange (solid) and red (dashed) lines represent the PDF of the angle the velocity forms with the local magnetic field at the initial and final time respectively. The green (dotted) and blue (solid-dotted) ones represent the angle that particles velocity forms with the main magnetic field in the z direction at the same two instants. Both the two plasma scenarios start with an isotropic distribution of the velocities with respect to the magnetic field. Then the less magnetized particles(low \pb plasma) become strongly aligned with the magnetic field. The same effect is evident, though much less pronounced, for less magnetized particles (low \pb plasma).}
\label{fig:cosine}
\end{figure}

This acceleration mechanism in the out-of-plane direction has a global effect of elongating the ion velocity distribution function (VDF). In order to see if this typical alignment effect is present in our numerical experiments, we computed both the PDF of the angle between particles velocity and the local magnetic field and the PDF of the angle the particles velocity has with the main magnetic field (along z)
\begin{equation}
\cos({\theta}) = \frac{{\bm v}\cdot{\bm B}}{|{\bm v}||{\bm B}|}, \qquad \cos(\psi) = \frac{v_z}{|\textbf{v}|}
\label{eq:cos}
\end{equation}
In \fig{fig:cosine} we report the distributions of both the angles, at the initial and final times of the simulation, for different values of $\beta$. At the initial time, when turbulence is very \quotes{young}, the distributions are quite flat, meaning particles are moving isotropically. As the simulation goes on, particles tend to align with the main magnetic field in the z direction. This effects is much more evident for the low \pb plasma, the more magnetized one. Another feature one can notice in the low \pb plasma is that the particles orient themselves more on the local magnetic field rather than its z component, although the z component is its main one and the difference is not statistically relevant.

\begin{figure}
\centering
\subfloat
{ \includegraphics[width=\textwidth]{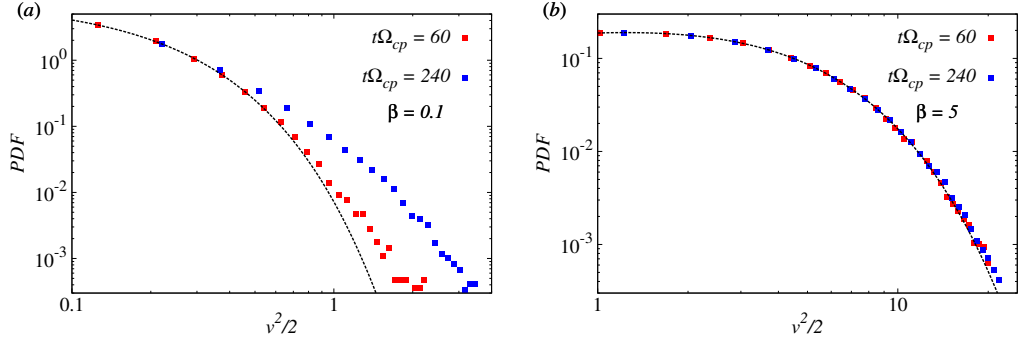} }
\caption{PDF of the particles energy, for the high and low \pb simulations, at different times of the simulation. The high \pb distribution does not change during the simulation, whereas the low \pb particles have a substantial energy gain because of turbulence.}
\label{fig:PDFE}
\end{figure}

It is crucial to see now whether this acceleration mechanism, locally related to intense parallel electric fields, can actually increase the particles energy, and up to which values. This correspondence is not trivial since the particles with anomalous acceleration are only a small fraction of the plasma. We have constructed the energy PDF to see whether the energy has a different behavior at different $\beta$. The energy PDFs for the low and high \pb values are show in \fig{fig:PDFE}. In the low \pb scenario, the PDF develops a power law tail, already seen in observation and previous numerical simulations. Whereas in the high \pb plasma the energy remains similar to the initial distribution, namely close to Maxwellian distribution. This suggests that the acceleration mechanism can energize particles, and that the process depends on the plasma $\beta$.

\subsection{Magnetic Trapping}

\begin{figure}
\centering
\subfloat
{ \includegraphics[width=\textwidth]{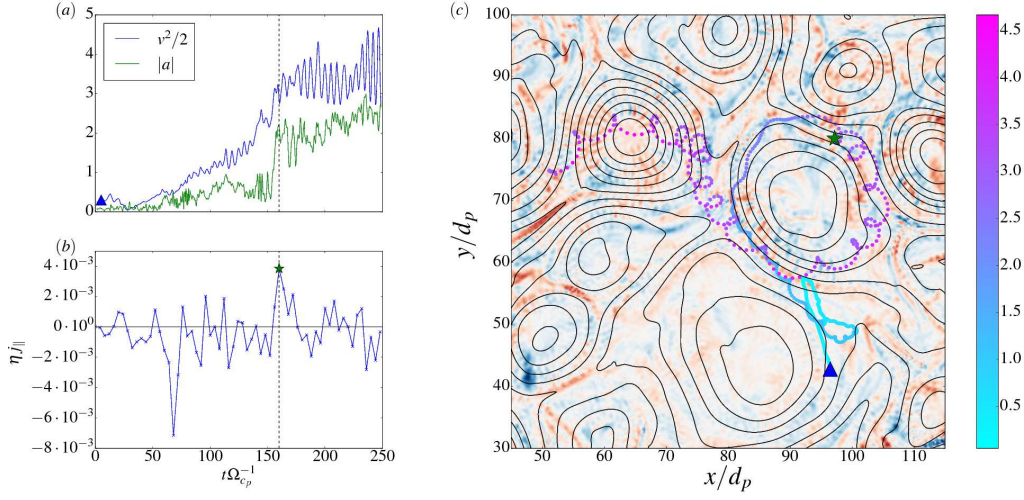} }
\caption{One of the most energetic particles followed along its path. Panel \pan{a} shows the particle energy and its acceleration. All of them show an increase in correspondence of a peak in the parallel current density reported in \pan{b}. Panel \pan{c} shows the colour map of the parallel electric field with the particle path which colour represents the particle's energy. The colour clearly shows that the particle gets energized when it remains trapped in the magnetic island until it gets enough energy to escape. The blue triangle represents the starting point of the trajectory and the green star is the point where the maximum value of parallel current density is found.}
\label{fig:wonder}
\end{figure}

We have seen that the acceleration mechanism involves the parallel electric field. However, there must be something else that makes this field energizing particles in low \pb plasmas. In figure \fig{fig:aQmap} we have seen that the anomalous particles lie within (or in between) magnetic islands, in agreement with previous discussions. The acceleration mechanism, evidently occurs in association with magnetic reconnection. Particles temporarily trapped inside small flux tubes or in the dynamically active region of larger flux tubes experience Fermi-like processes. In the  region near current sheets, particles can experience nearly continuous first order energisation \citep{Hoshino01,DrakeEA10,HaynesEA14}. In particular we followed the trajectory of one of the most energetic particles and monitored its energy, its acceleration, the energy derivative with respect to time and the parallel current density it samples during its journey (the parallel pressure term is not shown because it is too noisy for non-statistical treatment) (\fig{fig:wonder}). The figure also shows that the particle remains trapped in a magnetic island and its energy grows until it reaches a high enough energy to escape.

To estimate the escaping times we calculated the Lagrangian auto-correlation time using \eq{eq:corr_time}. This quantity represents the time at which each particle experience a correlated field, and can be estimated as
\begin{equation}
T_{esc} = \dfrac{1}{\langle v_x(t_0)^2 \rangle_a} \int_0^{\infty} \langle v_x(t_0) v_x(t_0+\tau) \rangle_a d\tau.
\label{eq:corr_time}
\end{equation}
In the above definition, the operation $\langle \cdot \rangle_a$ is the average only over the particles that have high acceleration. For the usual isotropy hypothesis. we computed the above relation along all directions, finding similar results. We obtained the following escaping times: $T_{esc}^{anom}(\beta=5)\sim 0.8 \Omega_{cp}^{-1}$, $T_{esc}^{anom}(\beta=0.5)\sim 3.0 \Omega_{cp}^{-1}$ , and $T_{esc}^{anom}(\beta=0.1)\sim 5.7 \Omega_{cp}^{-1}$. In addition we computed the same quantity for \textit{all} the particles, obtaining these values: $T_{esc}^{all}(\beta=5) = 0.95\Omega_{cp}^{-1}$, $T_{esc}^{all}(\beta=0.5) = 4.13\Omega_{cp}^{-1}$, $T_{esc}^{all}(\beta=0.1) = 9.64\Omega_{cp}^{-1}$.
This scenario is consistent with the fact that particles in low \pb plasmas develop high energy tails, since they are confined within magnetic islands for longer periods, experiencing the same parallel electric field. On the other hand, particles in high \pb plasmas easily escape from magnetic islands and are not efficiently energized. The comparison of the trapping times values at the same $\beta$ for the different kind of particle populations also supports this view. The escaping times calculated over all the particles are bigger than those of the anomalous particles suggesting that more energetic particles are more likely to escape from magnetic islands while lower energy particles remain trapped for longer period and can be coherently energised. This Fermi-like process, invoked in small scale reconnection \citep{Ambrosiano88}, is now quantitatively observed in large scale plasma turbulence. It is interesting now to further characterize the energisation process by looking at the characteristic parameters of these anomalous particles.

\subsection{The Magnetic Moment}

We now further inspect the acceleration process by looking at the magnetic moment $\mu$ of ions, defined as
\begin{equation}
\mu = \dfrac{m v_\perp^2}{2B}. 
\label{eq:mu}
\end{equation}

\begin{figure}
\centering
\subfloat
{ \includegraphics[width=\textwidth]{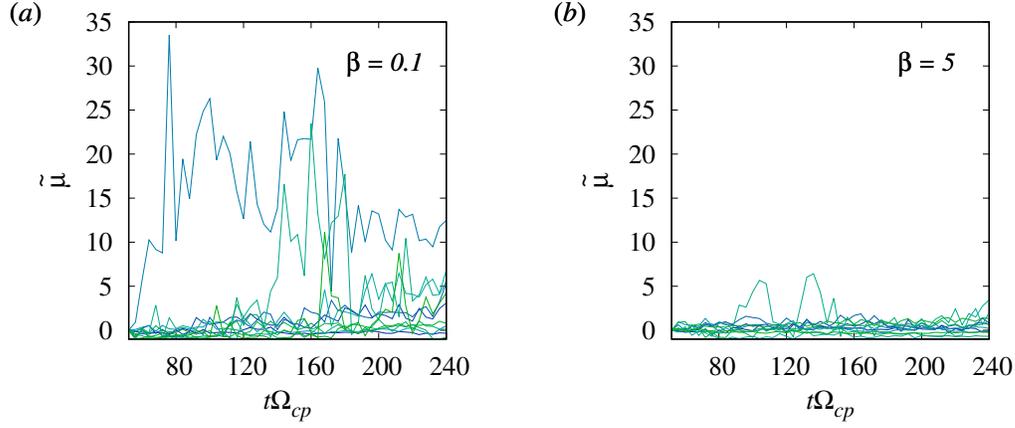} }
\caption{Magnetic moment as a function of time, for a group of 11 particles (same as in \fig{fig:vtraj}), for two values of $\beta$. In high-\pb plasma \pan{b}, the magnetic moment is much more constant (conserved).}
\label{fig:11murt}
\end{figure}

In the above expression, $m$ is the particle mass and $v_\perp$ its velocity (perpendicular to the magnetic field ${\bm B}$, measured at the particle's position). $\mu$ is an adiabatic invariant of the system, if $B$ is slowly varying. Indeed, the orbits are like closed circles and the flux of magnetic field passing through them is almost constant. This suggests that the magnetic moment might not be a constant of the motion in a turbulent system where several spatial scales are present, and where  magnetic field variations are neither negligible nor adiabatic \citep{Dalena12}. To quantify the behaviour of the magnetic moment for each particle, we computed the normalized moment
\begin{equation}
\tilde{\mu}_p = \dfrac{\mu_p(t)-\mu_p(0)}{\mu_p(0)}, 
\end{equation}
where the label $p$ now indicates a single particle. This measure gives us information about the variation of the particle magnetic moment with respect to its initial value. \fig{fig:11murt} shows $\tilde{\mu}_p$ as a function of time, for some particles, randomly selected, for two plasma \pb values. Particles that moves in low \pb plasma have the highest magnetic moment excursions, while in the high \pb plasma, where particles are  not so energized in time, their magnetic moment is much more conserved. It is important to notice that the magnetic moment distribution is very similar to the distribution of energies and acceleration, further confirming the relevance of this quantity for the process of plasma acceleration in turbulence.

\begin{figure}
\centering
\includegraphics[width=\textwidth]{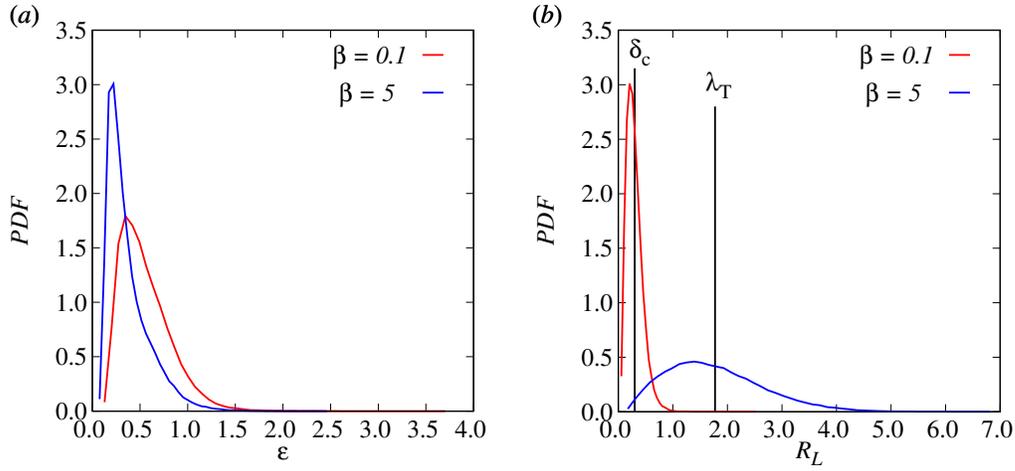}
\caption{\pan{a} PDF of the dispersion of the magnetic moment defined in \eq{eq:epsilonmu}. These PDFs show that particles are more likely to ``break'' their magnetic moment in the low \pb case. \pan{b} PDF of the Larmor radius for different $\beta$'s. High $\beta$ particles have a wide distribution, reaching values up to one order of magnitude bigger than in the low-$\beta$ case. The Taylor length $\lambda_T$ and the current width $\delta_c$ are also indicated. The Taylor length indicates the length of the biggest current sheet in the plane and sets an upper limit to the particles that can effectively get energized by interacting with the current sheets. Whereas the current width $\delta_c$ is the average current sheet thickness.}
\label{fig:epsilonmuRLPDF}
\end{figure}

To see whether the magnetic moment is a constant of the motion, in a more quantitative way (\fig{fig:11murt} refers only to a small portion of particles), we calculated the standard deviation $\sigma_\mu(p)$. This can be interpreted as the dispersion of the magnetic moment, for each particle, and is defined as 
\begin{equation}
\sigma_p = \sqrt{ 
\frac{1}{T} \int_{t_0}^{t_0+T} \left[ \mu_p(t')-\langle \mu_p\rangle_T \right]^2 dt'
}, 
\end{equation}
where $\mu_p(t)$ is the magnetic moment of the \ita{p}-th particle at the time $t$, and the average in the integral is calculated over the whole time. In case of a perfectly conserved magnetic moment, this quantity is null.
We have then built the PDF of
\begin{equation}
\varepsilon = \dfrac{\sigma_p}{\langle \mu_p \rangle_t},
\label{eq:epsilonmu}
\end{equation}
that indicates how much the magnetic moment deviates from its mean value -- how much the magnetic moment is ``broken''. The PDF$(\varepsilon)$ is shown in panel \pan{a} of \fig{fig:epsilonmuRLPDF}. As expected, the violation of the magnetic moment is much more pronounced (broader distribution) in the case of $\beta=0.1$. Low-\pb particles have small gyro-radii and they can probably interact with the local strong inhomogeneities.

We can relate the magnetic moment violation directly to the turbulence characteristic scales. Our analysis suggests that particles can interact with current sheets, if their Larmor radius is much shorter than the Taylor length $\lambda_T$ -- the largest current sheet size (\eq{eq:lambdat} ). If the radius is bigger, the particle can gyrate without even ``noticing'' the current sheet. The other important scale is the current width $\delta_c$, that is the mean size of the current sheet cores. Panel \pan{b} of \fig{fig:epsilonmuRLPDF} shows a more quantitative view of the above speculation, suggesting a kind of spatial resonance. In this figure we report the PDF of the Larmor radius along with the turbulence characteristic lengths $\lambda_{_T}$ and $\delta_c$.  The Larmor radii of particles in high \pb plasma are broadly distributed and are much bigger than the current sheets thickness (while they are on the order of the Taylor scale). The case with $\beta=0.1$ shows, instead, that the average $R_L$ are smaller than (or on the order of) $\delta_c$. In this scenario, particles with the resonant Larmor radius feel the presence of the sharp discontinuities, and undergoes a magnetic moment break. This finally leads to the acceleration mechanism, and to the energisation of ions in turbulent plasmas.

\subsection{Approaching to Velocity Space Diffusion}
\begin{figure}
\centering
\subfloat
{ \includegraphics[width=\textwidth]{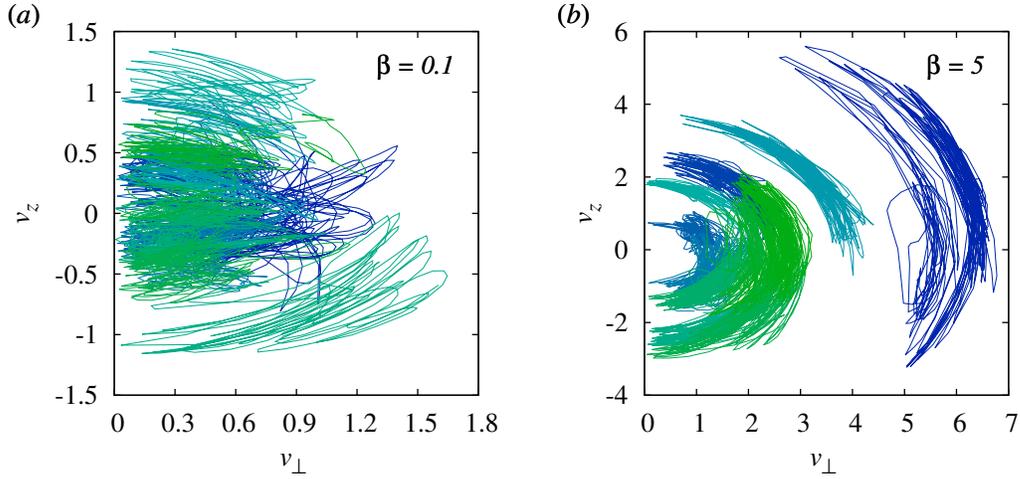} }
\caption{Velocity space trajectories for the particles of \fig{fig:11murt}, at high and low $\beta$'s. Particles moving in the high \pb plasma are more energetic and do not undergo energisation processes. They diffuse in pitch angle and hence move on quasi-isoenergetic shells. Differently, in the low \pb scenario, particles manifest more complex (ergodic) trajectories: they are not locked on isoenergetic shells  as they can effectively gain/lose energy throughout their magnetic islands journey.}
\label{fig:vtraj}
\end{figure}

We have studied the process of particles acceleration and energisation. Acceleration can be interpreted as a ``motion'' in the velocity space, i.e. $v$-diffusion \citep{Subedi17}. In this last section we will  briefly mention this aspect of diffusion. As for spatial diffusion, it is instructive  to observe particles trajectories in the velocity space. In particular, we represent these trajectories in a 2D space made by $v_z$ (parallel to the global magnetic field) and the in-plane velocity $v_\perp = \sqrt{v_x^2+v_y^2}$. The trajectories in the velocity space are shown in \fig{fig:vtraj}.

The velocity space trajectories show, from another point of view, the same behavior we have seen in the previous sections. In the high \pb plasma, particles are accelerated but not effectively energized by the parallel electric field. This means that they can diffuse only in pitch-angle and gyrophase, namely they move on the surface of iso-energetic shells, varying only the angle the velocity forms with the magnetic field. In the low \pb scenario, particles, instead, are accelerated and energized by the current sheets. The combined acceleration and energisation phenomena make particles trajectories in the velocity space more complex, since particles can diffuse both in pitch-angle and momentum-amplitude. In this case, ions change their velocity modulus and can acquire a continuous range of values making the trajectories almost ergodic. The ergodicity domain is bounded by the upper limit of energy a particle can gain before escaping a vortex and run into a decelerating (de-energizing) electric field. This interesting aspect of diffusion will be further inspected in future works.

\section*{Discussion and Conclusions}
In this work we described the diffusion and the acceleration of ions in plasma turbulence, using self-consistent kinetic simulations (kinetic ions and fluid electrons). These topics are of primary importance both in astrophysical and laboratory plasmas. We simplified the problem by using a 2D geometry, which can be a valid approximation to understand the nature of strongly anisotropic (magnetized) fluctuations \citep{Shebalin83,Dmitruk04,Matthaeus86}. Particles have been evolved making use of a PIC algorithm. The PIC algorithm is very useful and of practical fundamental importance when treating non-equilibrium plasmas for which the wave-particle interactions are crucial. The 2D hybrid-PIC simulations have been performed for three different $\beta$'s, in order to reproduce a wide range of physical systems. The different values of $\beta$ used vary for more than one order of magnitude allowing us to describe plasma scenarios spanning from the solar corona to fusion devices. The 2D maps of the current density along with the magnetic field showed the development of fundamental turbulent structures, such as vortices and current sheets.

We have then studied the motion of ions, moving self-consistently in the electromagnetic field. As suggested by previous (and numerous) test-particles studies, the motion is very erratic. Particles can be trapped in magnetic vortices or  scattered away by current sheets, wandering like a pollen in the atmosphere \citep{Servidio16}, or field lines in the solar corona \citep{Rappazzo17}. This kind of trajectories are achieved after rather long time intervals and can be statistically described within the theory of diffusion. The particle motion becomes uncorrelated when the particle is no longer trapped by the same vortex. Low energy particles have been found to have longer correlation times as they cannot easily escape from vortices.

All the existing theories on the diffusion coefficient can describe 3D systems and all these theories have been verified only with test-particle models. The NLGC theory is possibly the most precise theory that gives an estimate of the diffusion coefficient. We have then \quotes{reduced} this theory to the 2D case. This 2D NLGC theory has been found to be valuable in describing the diffusion coefficient of particles moving in self-consistent turbulent fields.

From the acceleration process, the PDFs show that the acceleration nature depends on the plasma $\beta$. Acceleration is a stochastic variable for high \pb plasmas, whereas it is distributed with power-law tails for low \pb case. We have found that the electric field component parallel to the magnetic field is correlated with particle acceleration. The ``anomalously'' accelerated particles, i.e. the particles with acceleration values that exceed the variance of the distribution, are connected to regions with high parallel electric field. This has been seen, qualitatively, by spotting these particle on the parallel electric field map and, quantitatively, by computing the conditional statistics.

By looking at the energy PDFs, we noticed that particles in low \pb systems, such as in the solar wind, are more effectively energized. This process can be linked to the presence of narrow current layers. The main phenomenon acting on the particle is the breaking of the particle magnetic moment. Local spatial resonances break this constant of motion, leading to acceleration and finally to energisation of particles. In this process, particle undergo a spatial resonance with the background turbulent structures: ions that have their Larmor radius on the order of the current sheet thickness experience large excursions of their magnetic moment. These particles experience a local acceleration process, while in the case with much larger $\beta$, as can be found in the magnetospheric environment, the plasma elements do not ``see'' the reconnecting current sheets and the embedded parallel electric field. This kind of interaction with vortices and current sheets has been proved also by looking at characteristic scales resonances. Particles characteristic scale is the Larmor radius, whereas for turbulence we computed the Taylor length and the current width. A consistent percentage of high energy particles has a Larmor radius bigger than the Taylor length, that is, qualitatively, the in-plane length of the  biggest current sheet. This means high energy particles can barely ``notice'' the current sheets. 

Finally, we have introduced the concept of velocity space diffusion, because above processes might be related to the stochastic motion of particles in the velocity space. We still lack a fundamental theory to determine the diffusion coefficient in velocity space \citep{Miller90,Miller95}, and we leave this work for future studies. Future works can be focused on an analytical treatment of the velocity-space diffusion. Moreover, a full 3D study will be performed in the future, taking into account also the role of kinetic electrons.

\section*{Acknowledgments}
This work is partly supported by the International Space Science Institute (ISSI) in the framework of International Team 405 entitled “Current Sheets, Turbulence, Structures and Particle Acceleration in the Heliosphere", by the US NSF AGS-1156094 (SHINE), and by NASA
grant NNX14AI63G (Heliophysics Grandchallenge Theory), the MMS mission through grant NNX14AC39G, and the Solar Probe Plus science team (ISOIS/Princeton subcontract SUB0000165).

\bibliographystyle{jpp}

\bibliography{biblio}

\end{document}